# Generation and Classification of Activity Sequences for Spatiotemporal Modeling of Human Populations

Albert M Lund,[1] Ramkiran Gouripeddi [1,2] and Julio C Facelli[1, 2, *]

[1] Department of Biomedical Informatics and [2] Center for Clinical and Translational Science,

The University of Utah, Salt Lake City, Utah 84108

**Abstract**

Human activity encompasses a series of complex spatiotemporal processes that are difficult to model, but represents an essential component of human exposure assessment. A significant empirical data source like the American Time Use Survey (ATUS) can be leveraged to model human activity, but tractable models require a better stratification of activity data to inform about different, but classifiable groups of individuals that exhibit similar activities and mobility patterns. We have developed a simple unsupervised classification and sequence generation method from existing machine learning algorithms that is capable of generating coherent and stochastic sequences of activity from the data in the ATUS. This classification, when combined with any spatiotemporal exposure profile, allows the development of stochastic models of exposure patterns for groups of individuals exhibiting similar activity behaviors.

**Keywords:** American Time Use Survey, Machine Learning, Random Forests, Classification, Exposure Modeling

[*] Corresponding author at:
Department of Biomedical Informatics
421 Wakara Way, Suite 140
Salt Lake City, Utah  84108
Phone: 801-581-4080
E-mail:  Julio.Facelli@utah.edu

**Introduction**

Estimating human exposure to airborne and other broadly distributed pollutants presents a major challenge, because humans are mobile and inhabit a variety of microenvironments. It is insufficient to model only the spatiotemporal distribution of pollutants, as even for a geographically homogeneous distribution of pollutant different individuals will experience different levels of total exposure depending on their activity patterns (1-3). Any successful model of human exposure requires an estimation of the activities human agents perform, along with their locations and context of those activities. Furthermore, a large amount of pollutant emissions in urban environments results directly from human activity, therefore modeling human activity has also potential use in estimating directly pollution distributions from mobile sources, like automobile emissions.

Comprehensive and detailed activity patterns from individuals can be gathered using a variety of tracking devices, but such methods may be cumbersome to implement, prone to privacy concerns, and may fail to capture contextual data (4). On the other hand the American Time Use Survey (ATUS) (5) provides a comprehensive picture of human activities in the United States of America (US) and can be used to infer human behavioral patterns. The ATUS dataset is highly complex, with each annual survey containing over 10,000 activity diaries of recorded temporal sequences detailing the daily activities of the survey respondents. Each activity diary can include up to 80 discrete activities with dozens of auxiliary variables and additional demographic variables embedded in the dataset. The composition and timing of activities can have significant overlap, but also present distinct patterns based on demographics. For example, we expect that the majority of the respondents will report sleeping, eating, and grooming in most activity diaries, but other activities, such as working, recreation, and child care, will have unequal representation across different demographic categories. The ATUS has been described in great detail, including a comprehensive descriptive analytics analysis in a recent publication (6).

The degree of complexity in the ATUS makes expert analysis or the development of a gold standard difficult; its dimensionality and size are above the threshold for effective analysis or visualization. The synthesis of activity sequences has been explored with varying levels of success (7-11), but to the authors' knowledge, no attempts to classify activities from the ATUS for cohort identification have been reported. Classification of individual activity patterns is a critical step for the development of stochastic models of exposure (12). To address this need we developed a method for unsupervised classification of the ATUS data that broadly classifies activity and demographics without relying on human expertise. Identification and classification of activity and demographic classes enables us to construct activity sequences, which are artificial constructs used to model behavior in our agent based model (13). We developed a simple approach to construct activity sequences using the concept of starting windows – which are periods where an activity may start. We then show that our method of generating activities results in sequences that are qualitatively indistinguishable from those collected in the ATUS.

**Methods**

*Classification of the ATUS Activity Diaries*

While intuitively, we can conceive that different individuals follow different activity patters, to our knowledge, there are not studies that have organized these activities in a formal way and common patterns have been found, such that individuals can be classified according to them. The ATUS activity diaries are organized into multiple tables containing demographic properties of the respondents (age, gender, work status, married status, etc.), activities for each respondent (a sequence of records containing activity type, start times, length), and some auxiliary information describing household composition and activity context. Variables can be categorical or continuous, possibly censored to protect unique respondents, and have hierarchal dependencies based on survey responses. We eliminated variables from the demographic table relating to survey questions that had low response rate and/or low variance, as these would be non-informative. Our final selection contains 12 demographic variables listed in Table 1, all of which can also be inferred from the US Census and employment statistics.

We transformed the activity tables into two separate vectors representing the activities reported by each individual participating in the survey. The first vector with approximately 400 dimensions counts the number of instances that each activity found in the activity diary of each individual. The second vector with 288 dimensions discretizes the 24-hour period of each individual's diary into five-minute intervals, assigning the code of the primary activity reported in each slice to the corresponding slot. Together, these activity vectors capture both the categorical and temporal pattern of activities for each respondent. We used these vectors along with the 12 demographic variables to create the feature set for activity classification (Table 1).

Our approach to classifying activities and demographics was as follows. First, we generated a random forest with 2000 truncated trees having a maximum tree depth of five leaf nodes. We used the Random Trees Embedding method from scikit-learn to generate this forest, which generates a random forest on random subdivisions of variables in the absence of labels (14). We then generated a proximity matrix according to the method proposed by Breiman (15), by counting the number of times each pair of feature vectors appear on the same leaf node for each tree in the initial random forest. This proximity matrix is used as the input for a two component t-

Stochastic Neighbor Embedding (t-SNE) (16), which is used for embedding high-dimensional datasets in low dimensional spaces. Next, we normalized the embedded coordinates from t-SNE to the interval (-1,1) and performed clustering using density-based spatial clustering of applications with noise (DBSCAN) (17). We manually estimated the maximum cluster distance and sample parameters, since these hyperparameters are dependent on the dataset and features used. We used a maximum cluster distance values of 0.03 for a cluster size of 20 and 0.02 for a cluster size of 10 for the demographic and activity feature sets, respectively. Using these parameters allowed us to strongly select small dense clusters and for feature vectors to be non-labeled by the algorithm.

The product of the DBSCAN clustering generates a set of labeled and unlabeled feature vectors. We use the labeled feature vectors to train a truncated Extra Random Forest (18). The maximum tree depth of this forest is eight, using the entropy criterion (which is preferred for categorical data). We then classify all unlabeled feature vectors using this new random forest. We do this because the initial clustering leaves up to 30% of feature vectors unlabeled, and many of the labeled features are similar enough to be classified the same. As we needed our classes to have some level of statistical power, we generated one additional set of random forests, using the same parameters, but this time without truncation (no maximum tree depth). This set of forests was trained on all classes above a size cutoff of 25 feature vectors, with the remaining small classes being classified according to this new classifier. This produces the final classes for the demographic and activity classes, and generates a classifier that can be used in conjunction with the US Census as part of our agent based model described in the next article of this issue (13).

*Generation of Activity Sequences using Starting Windows*

While the classification by itself is a useful tool for identifying distinct patterns of activity, it is insufficient for predicting or simulating the behavior of an arbitrary agent representing a person. The activity classes generated by our classifier provides a basis for what patterns of activity exist, but the activity diaries themselves are not suitable for simulation purposes because they are intrinsically tied to the empirical and geographical constraints on the persons interviewed for the ATUS. Instead, we generate synthetic activity sequences from a probabilistic representation of each activity class.

We generated synthetic activity sequences for each activity class according the following procedure. For each class, we considered each activity present in the cohort separately and collected their starting times. We defined the concept of a starting window - which is a period of time when an activity may start. Using Bayesian Gaussian Mixtures (19) we generated a set of one dimensional clusters of activity starting times to create starting windows, which define a period of time when an activity can start. For example, if we were to distinguish daytime naps and nighttime sleeping, we would define two separate starting windows for each type of activity based on starting time, even though both instances are classified as sleeping activities

Utilizing these starting windows, we calculated four different probabilities. First, we calculated the probability that a member of the activity cohort will perform an activity defined by a starting window. This is the probability of a starting window appearing in an arbitrary sequence drawn from the set of activity diaries that contains the starting window of interest. This probability captures the idea that some activities are repeatedly and consistently performed across the population, such as sleeping, eating, and personal grooming, but also allows for exceptions in common behavior. We expected the members of each activity class to follow a schedule, but with potential variations. The second probability we calculated is the joint probability between start windows and activity lengths. We cluster activity lengths into length windows that are generated the same way as start windows, but using activity lengths instead of start times. The reason for using length windows instead of a more common distribution is that activity lengths can exhibit very different scales depending on context. For example, a nap could last anywhere from twenty minutes to three hours long, whereas a typical night sleep might vary from four the twelve hours. Further, activity lengths can have unusual distributions and cluster in ways that do not approximate to a smooth function. The third probability we calculated is the probability that an activity in one start window is preceded by an activity in another start window. This captures the idea that the order of some activities can be indiscriminate or based on preference, while others have specific causal orders. For example, food preparation always precedes the actual activity of eating, but the order of reading a book and watching a movie for evening entertainment largely depends on the preference of the participant. Estimating this probability allows us to effectively sort activities, and insert the necessary stochastic components needed to capture variability in activity order. Finally, we calculated the joint probability between start window and location type. Although the ATUS does not have specific geographic locations in the dataset, it does define the type of location for each activity (e.g., home, workplace, store, etc.). Encoding these location types allows us to utilize contextual information for assigning concrete locations to activities in a synthetic activity sequence.

We utilize these four probabilities to generate synthetic activity sequences using Monte Carlo sampling. For this, we selected a set of start windows, assigned activity lengths, sorted those starting windows stochastically, and then assigned locations types. Next, we inserted travel activities between activities that occur at different locations to improve the quality of the sequence. Finally, we adjusted activity lengths within the tolerance of the starting windows and minimum or maximum activity lengths to fill the period of simulation so that there are no gaps in the synthetic sequence. We performed this adjustment using a weighted coefficient based on the selected length of each activity to preserve the relative lengths of activities.

**Results & Discussion**

Figure 1 shows example activity classes derived from the classification process. Distinctive patterns of activity can be isolated despite the simplicity of the classification algorithm. Significant overlap in activity profiles occur between some classes, especially in classes where the fundamental activity profiles are essentially the same, but the timing of activities can be shifted as in cases where shift work is represented. This suggests that the classification method is effective in making distinctions in both temporal and categorical domains. For these experiments, the demographic classification produced 95 classes with a median class size of 83 records and maximum of 696, while the activity classification produced 76 classes with a median class size of 82 records and maximum of 1237; both classifiers have an artificial minimum of 25 records. The number of classes produced by this approach varies due to stochastic elements in the t-SNE and random forest algorithms. We attempted to broadly classify the activity classes based on the main category of non-sleep activity that dominated each activity record. Roughly, 40% of classes are dominated by work activities, while 25% are dominated by recreational activities. The remaining 35% of classes comprise some mixture of household activities, child or elderly care, and school related activities.

Figure 2 shows sets of real activity sequences from the ATUS and synthetically generated activity sequences for a typical day belonging to a member of the working class. Qualitatively the two sets are difficult to distinguish from each other. Distinctive temporal boundaries are present between some activities in the real sequences, which are an artifact of the classification algorithm strongly selecting a subset of temporal features. These temporal boundaries disappear in the synthetic activity sequences due to the length adjustment step and the introduction of randomness from the Monte Carlo process. Despite this variation, the overall profile of activity in the synthetic sequence still visually captures the overall prevalence of activities.

We performed a quantitative analysis on our synthetic sequences in order to validate their similarity to the real sequences. Because the temporal sequences are categorical in nature, a detailed temporal analysis of the synthetic sequences is difficult. A realistic way to compare categorical temporal sequences is through a binary comparison at the smallest temporal granularity. Groups of sequences can be compared through their statistical mode (most frequent observation) and a measure of dispersion like the Gini index (20) which are analogous to the mean and standard deviation of a normally distributed continuous variable. We calculated the modes of each activity class by determining the most frequent activity at each minute across all activity sequences in that class. We then made a binary comparison between the modes of the synthetic and the ATUS reported sequences to obtain a percentage similarity between the two. We obtained the Gini index by calculating the frequency of all activities for each minute across all activity probabilities. We compared the synthetic and reported sequences by performing a linear regression of the Gini index over time.

Figure 3 shows the plot of the r-correlation of the Gini indices and mode similarities for all activity sequences. The majority of activity classes (61%) have both Gini correlation and mode similarities above 0.8, while 95% of classes are above the 0.6 threshold. This presents a strong evidence that our sequence generation algorithm correctly reproduces the majority of the activity classes.

In the development of the sequence generation algorithm, we explored several techniques. Our results from using a simple Markov chain ended up being intractable with the generated sequences having little to no resemblance to the ATUS data and incapable of capturing the structured nature of some activities (especially the home-work-home pattern). We also tried to train a recurrent neural network (RNN) against the ATUS activity diaries, but we found that the activity sequences were too short to train reliably the RNN. Specifically, we believe that the RNN needed to be trained on activity sequences spanning multiple days, which are unavailable from the ATUS surveys, which only cover 24-hour periods. However, we ultimately found that the method we developed was both simpler and easier to implement than an RNN, and required less computational effort to establish and generate sequences. The method we have developed and presented here is also substantially more explainable compared to an RNN.

**Conclusions**

We successfully developed and demonstrated a generalizable method to classify human activity sequences and generate synthetic spatiotemporal activity sequences.  While in this study we derived activity sequences from the ATUS activity classes, our method  is not specific to this survey, and can used for any well-structured activity survey data sets. We believe that the application of this approach will enable researchers to make important inroads into simulating human activity patterns at population levels - a first step in generating comprehensive exposome records and their utilization in translational research.

**Acknowledgements**

The research reported in this publication was supported in part by NIBIB/NIH under Award Number 1U54EB021973 and NCATS/NIH under Award Number UL1TR001067. Computational resources were provided by the Utah Center for High Performance Computing, which has been partially funded by the NIH Shared Instrumentation Grant 1S10OD021644-01A1.

**Conflict of Interests**

The authors declare no competing financial interests in the publication of this work.

.

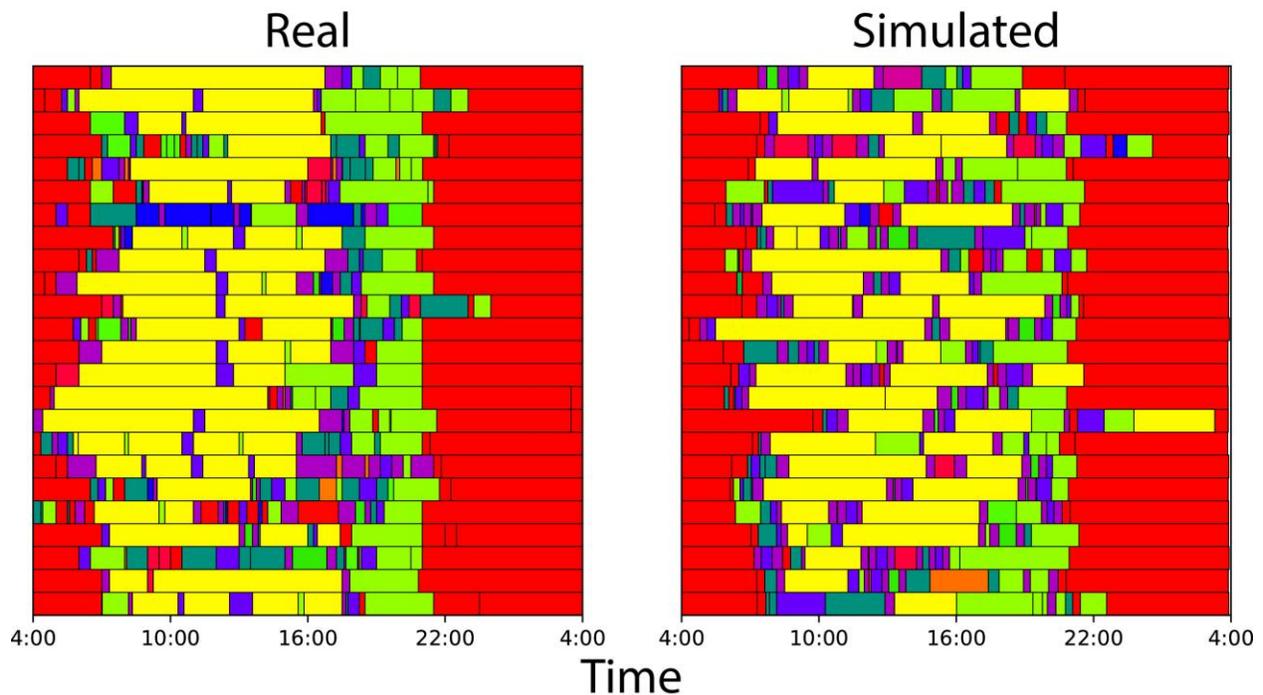

**Figure 1**. Examples of activity classes generated by the unsupervised classification method. Distinct patterns of activity can be identified from the method. Panel A depicts a cohort that primarily participates in recreation activities (watching TV, reading, attending events), while panel B depicts a cohort that mostly participates in household activities (cleaning, yard work, child care, etc.). Panels C and D depict two different shifts of working days. The fact that the algorithm can elucidate temporal patterns is especially useful.

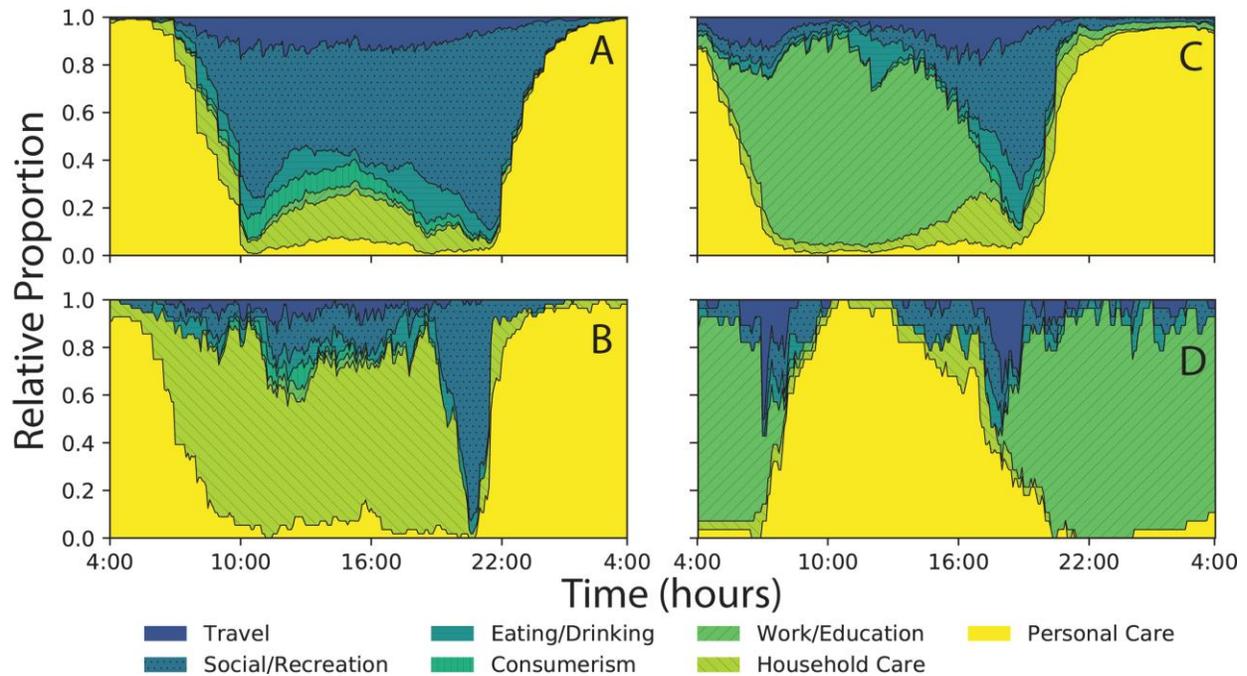

**Figure 2**. Real and simulated sequences for a single class, shown in their sequential form. Each row represents a different sequence, while different colors represent different classes of activities. Generally the simulated sequences conserve the same relative pattern of activity as the real sequences. Deviation from the strict timing of the real sequences is expected since the sequence generation algorithm includes some smearing components.

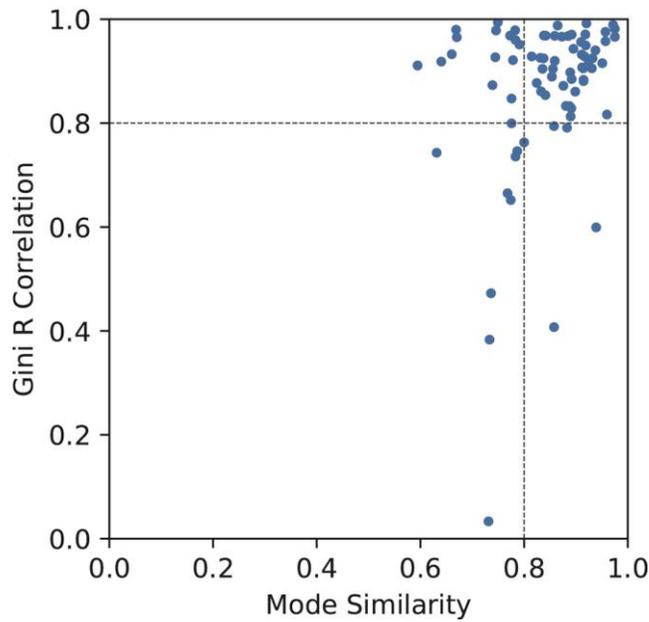

**Figure 3**. Similarity plot of synthetic and measured activity sequences. For each type of activity sequence, the most frequent activity (the mode) and Gini index is calculated for each minute across the cohort. The mode similarity is the fraction of minutes where the most frequent activity is the same between synthetic and measured sequences. The Gini R Correlation is from the linear regression of the Gini Indices for each minute. 61% of activity sequences have both similarities and r-values greater than 0.8.

**Table 1**. List of 12 demographic variables and activity vectors included in the data classification. Variable names are given as they appear in the ATUS and are relative to the survey respondent in each record. The demographic classifier uses only these 12 variables, while the activity classifier used the 12 variables and associated activity vectors as feature set.

|  | FEATURE NAME | DESCRIPTION |
| --- | --- | --- |
| **DEMOGRAPHIC VARIABLES** | TEAGE | Age |
|  | TEHRUSL1 | Hours worked at main job |
|  | TELFS | Labor force status (employed, unemployed, not in labor force) |
|  | TESCHENR | Enrolled in high school, college or university |
|  | TESCHFT | Enrolled as full time or part time student |
|  | TESCHLVL | School enrollment level (high school, college, or university) |
|  | TESEX | Gender |
|  | TESPEMPNOT | Employment status of spouse or unmarried partner |
|  | TESPUHRS | Hours worked by spouse or unmarried partner |
|  | TRCHILDNUM | Number of household children under age 18 |
|  | TRDPFTPT | Full time or part time employment status |
|  | TRHHCHILD | Presence of household children under age 18 |
|  | TRSPPRES | Presence of spouse or unmarried partner in household |
|  | TUDIS2 | Disability preventing work in the next six months |
|  | TUELNUM | Number of elderly people cared for this month |
|  | TUSPUSFT | Spouse or unmarried partner full time or part time employment status |
| **ACTIVITY VECTORS** | Activity count | The number of times each type of activity is performed in the activity diary. Contains approximately 400 activity counts |
|  | Activity Time | The main activity performed in each five minutes slice in each activity diary. There are 288 five minute slices in a single day. |